\documentclass[
reprint,
superscriptaddress,
amsmath,
amssymb,
aps,
longbibliography,
prb,
showpacs,
floatfix
]{revtex4-2}
\usepackage{hyperref}
\usepackage{graphicx}
\usepackage{xcolor} 
\usepackage{svg}
\usepackage{tabularx}
\usepackage{multirow}
\usepackage{dcolumn} 
\newcolumntype{d}[1]{D{.}{.}{#1}}
\newcolumntype{L}[1]{>{\raggedright\arraybackslash}p{#1}}
\newcolumntype{C}[1]{>{\centering\arraybackslash}p{#1}}
\newcolumntype{R}[1]{>{\raggedleft\arraybackslash}p{#1}}
\usepackage{amsmath}
\usepackage{bm}
\usepackage{amssymb}
\usepackage{stackengine} 

\usepackage{braket}

\usepackage{comment}
\usepackage{enumitem}

\usepackage{soul}
\definecolor{color1}{rgb}{0,0.25,0.70}

\hypersetup{colorlinks=true, 
    linkcolor={color1},
    citecolor={color1}, 
    urlcolor={color1}
}

\begin{document}

\title{Light-induced weak ferromagnetism through nonlinear magnonic rectification}

\author{Tom\ Kahana}
\email{tomkahana@mail.tau.ac.il}
\affiliation{School of Physics and Astronomy, Tel Aviv University, Tel Aviv 6997801, Israel}
\author{Daniel\ A.\ Bustamante Lopez}
\affiliation{Department of Physics, Boston University, Boston, Massachusetts 02215, USA}
\author{Dominik~M.\ Juraschek}
\email{djuraschek@tauex.tau.ac.il}
\affiliation{School of Physics and Astronomy, Tel Aviv University, Tel Aviv 6997801, Israel}

\date{\today}


\begin{abstract}

Rectification describes the generation of a quasistatic component from an oscillating field, such as an electric polarization in optical rectification, or a structural distortion in nonlinear phononic rectification. Here, we present a third fundamental process for magnetization, in which spin precession is rectified along the coordinates of a nonlinearly driven magnon mode in an antiferromagnet. We demonstrate theoretically that a quasistatic magnetization can be induced by transient spin canting in response to the coherent excitation of a chiral phonon mode that produces an effective magnetic field for the spins. This mechanism, which we call nonlinear magnonic rectification, is generally applicable to magnetic systems that exhibit infrared-active chiral phonon modes. Our results serve as an example of light-induced weak ferromagnetism and open a promising avenue towards creating dynamical spin configurations that are not accessible in equilibrium.

\end{abstract}

\maketitle


\section{Introduction}

\begin{figure}[b]
\centering
\includegraphics[scale=0.074]{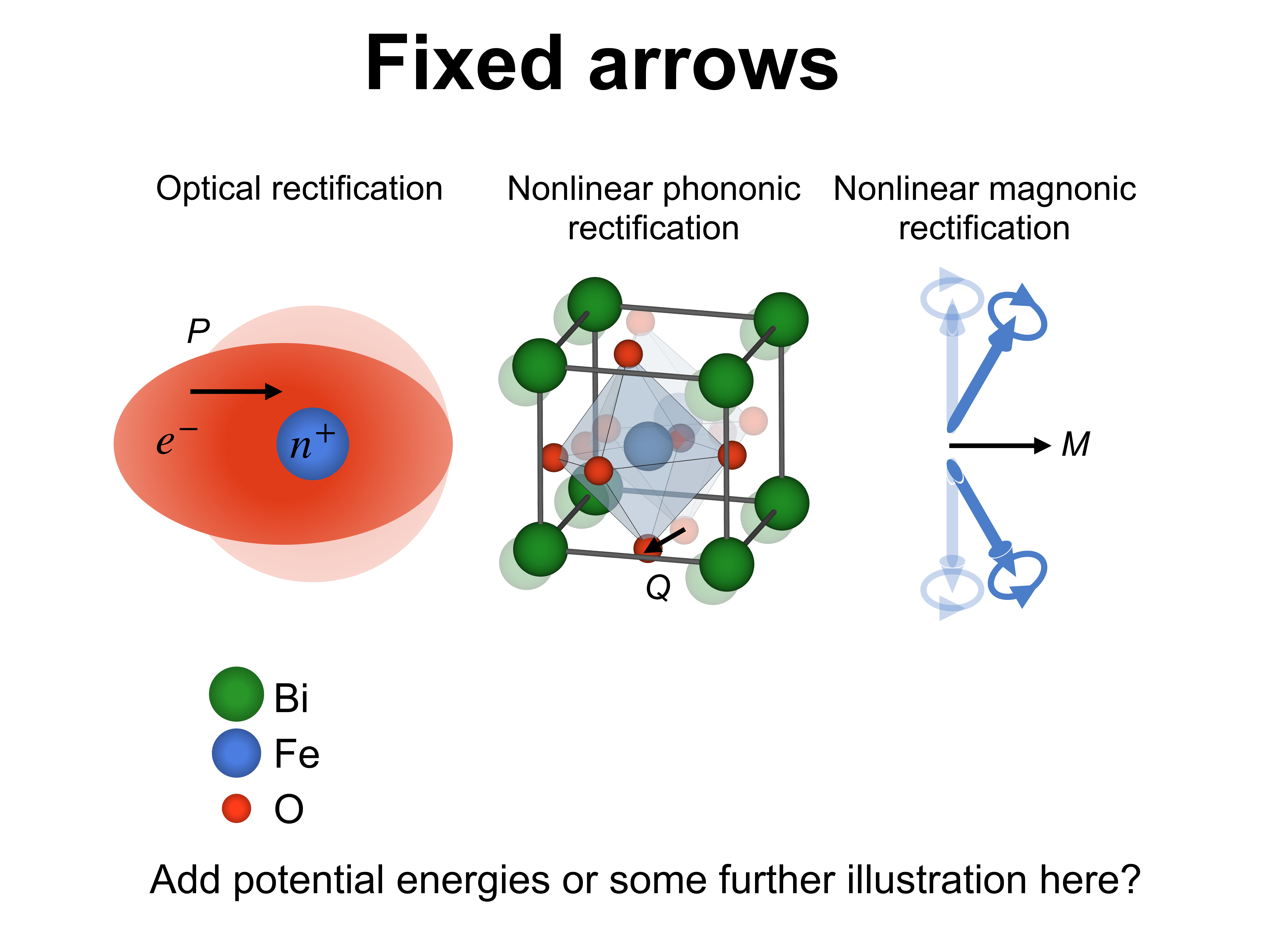}
\caption{Schematic overview of rectification processes in solids. (Left) Optical rectification induces an electric polarization, $P$, through a distortion of the electron cloud with respect to the nucleus \cite{BoydNonlinearOptics}. (Middle) Nonlinear phononic rectification induces a lattice distortion, $Q$, through a shift of the equilibrium value of vibrational atomic motion \cite{forst:2011}. (Right) Nonlinear magnonic rectification induces a magnetization, $M$, through a transient spin canting (this work).}
\label{fig:overview}
\end{figure}

Rectification is a powerful tool that enables the conversion of a time-dependent oscillating field into a quasistatic DC component. The most prominent example is optical rectification, in which a quasistatic electric polarization is generated from the oscillating electric field component of light in a dielectric material, as sketched in Fig.~\ref{fig:overview}, and which has become one of the most important mechanisms for the generation of terahertz radiation today \cite{Dhillon2017}. Within the last decade, a new rectification mechanism has been established for lattice vibrations that are excited coherently by ultrashort laser pulses, which is based on the intrinsic anharmonicity of the interatomic potential-energy surface. A resonantly driven phonon mode exerts a unidirectional force onto another phonon mode, which is coupled nonlinearly to it. The coupling potential can be described as a product of the amplitudes of the driven ($d$) and coupled ($c$) phonon modes, $V=Q_d^2 Q_c$. As a result, the crystal structure  distorts quasistatically along the eigenvectors of the coupled mode, proportionally to the mean-square amplitude of the driven mode, $\langle Q_c \rangle \propto \langle Q_d^2 \rangle$ \cite{forst:2011,subedi:2014,fechner:2016}. This effect can be seen as a phononic analog of optical rectification, as illustrated in Fig.~\ref{fig:overview} \cite{forst:2011}. The quasistatic distortion due to the rectified phonon mode can change the electronic interactions present in the system and therefore modify the properties of the solid. In recent years, the manipulation of various complex electronic phases has so been achieved, including superconductivity \cite{mankowsky:2014,Mankowsky:2015,fechner:2016,Liu2020}, ferroelectricity \cite{subedi:2015,Mankowsky_2:2017,Li2019,Nova2019,Abalmasov2020,Chen2022}, and magnetism \cite{Fechner2018,Khalsa2018,Gu2018,Radaelli2018,Rodriguez-Vega2020,Disa2020,Afanasiev2021,Rodriguez-Vega2022,Disa2023}.

With these recent developments as a background, our present study asks the question: Does an analog mechanism also exist for magnons, collective spin precessions in magnetic materials? Here, we demonstrate the rectification of spin precession along the coordinates of a nonlinearly driven magnon mode, which leads to the generation of a quasistatic magnetization. The rectification is achieved by utilizing the effective magnetic field produced by coherently excited chiral phonons, which has recently been predicted and measured in various materials \cite{nova:2017,juraschek2:2017,Juraschek2019,Juraschek2020_3,Geilhufe2021,Juraschek2022_giantphonomag,Xiong2022,Basini2022,Geilhufe2023,Davies2023,Luo2023}. The coupling potential is fundamentally given by the angular momentum coupling between the chiral phonon mode and the electron spins, $V=\mathbf{L}_{ph}\cdot\mathbf{S}$. As a result, the spins cant and the axis of spin precession transiently tilts along the eigenvectors of the coupled magnon mode, as shown in Fig.~\ref{fig:overview}, inducing a magnetization proportional to the angular momentum of the chiral phonon mode, $\langle M\rangle \propto \langle L_{ph}\rangle$. The direction of magnetization can be switched by the reversing the helicity of the laser pulse and therefore of the chiral phonon mode. This effect is a magnonic analog of both optical and nonlinear phononic rectification, and in our example it leads to the light-induced creation of a nonequilibrium spin configuration resembling that of weak ferromagnets \cite{Cheong2007}.


\section{Theory of nonlinear magnonic rectification}

Our demonstration of nonlinear magnonic rectification consists of three parts: First, the derivation of a quasistatic magnetization that scales proportionally to the phonon angular momentum. Second, the derivation of a magnetoelectric susceptibility that describes the response of the magnetization to the excitation of the chiral phonon mode by the electric field component of light. Third, the demonstration of light-induced weak ferromagnetism through transient spin canting as a result of the coupled spin-chiral phonon dynamics.

\subsection{Model for spin-chiral phonon coupling}

\begin{figure}[t]
\centering
\includegraphics[scale=0.0825]{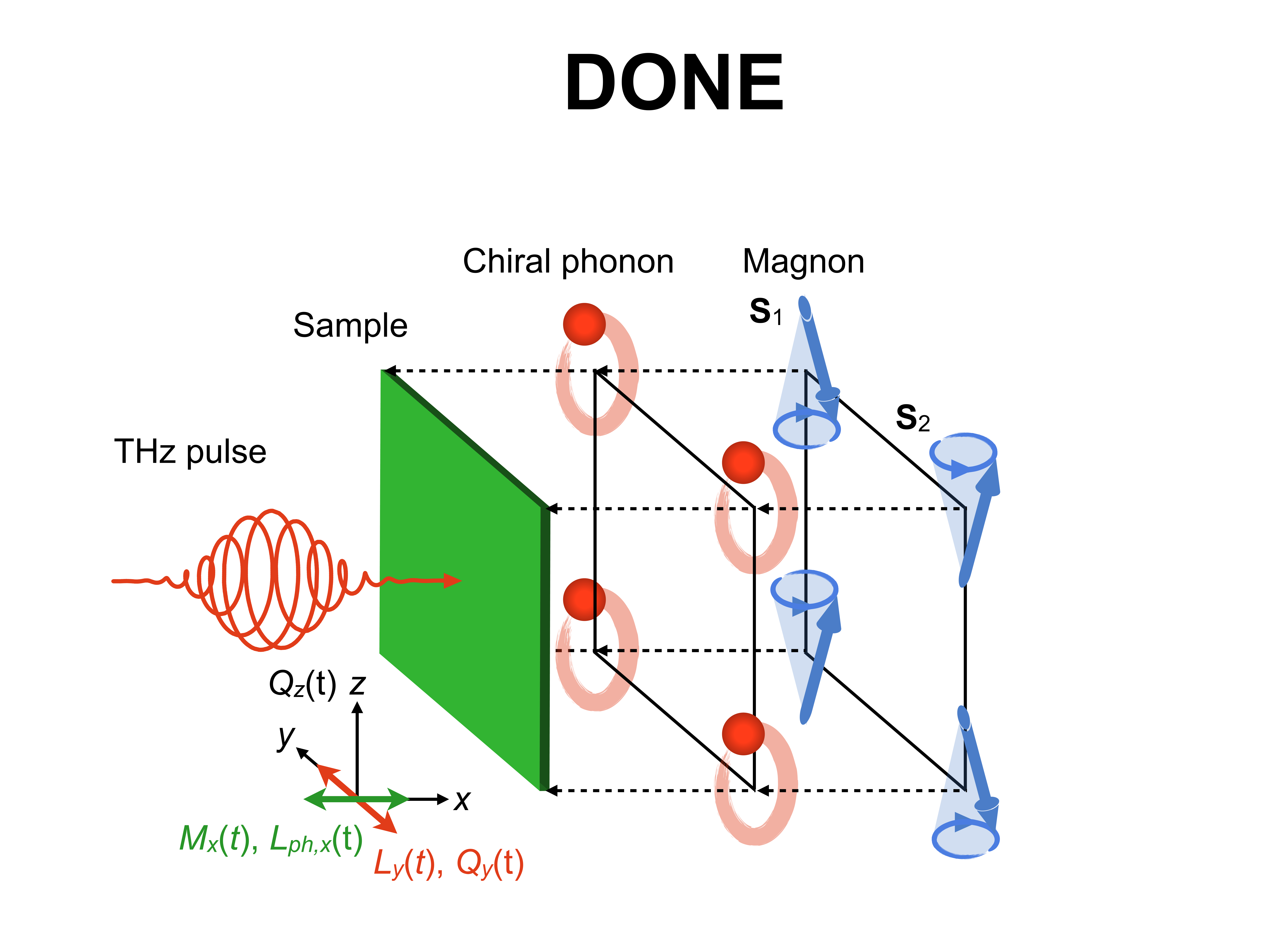}
\caption{Schematic of the antiferromagnetic system hosting a chiral phonon mode. We show the circular motions of the magnetic ions along the eigenvectors of the two components of the chiral phonon mode, $Q_y$ and $Q_z$, in the $yz$-plane. The phonon angular momentum, $L_{ph,x}$, accordingly points along the $x$-axis. We further show the spin precession along the eigenvectors of the high-frequency magnon mode, with the magnetization, $M_x$, pointing in the $x$-direction and the Ne\'{e}l vector, $L_y$, pointing in the $y$-direction.}
\label{fig:setup}
\end{figure}

We begin by developing the theory of nonlinear magnonic rectification in analogy to nonlinear phononic rectification. (We provide a prototypical example of nonlinear phononic rectification in Appendix~A.) While the dynamics of coherent spin precession in the macrospin approximation can be described by the Landau-Lifshitz-Gilbert equations \cite{Kampfrath2011,Rezende2019}, the dynamics of coherent phonon modes can be described by a phenomenological oscillator model \cite{subedi:2014,fechner:2016,Juraschek2018}. We use a combined approach as in Refs.~\cite{Fechner2018,Juraschek2021} to describe the coherent coupled dynamics of chiral phonons and magnons for the example of an antiferromagnetic Heisenberg model with easy-plane anisotropy, which provides a common representation of many antiferromagnetic materials \cite{Rezende2019}. For an antiferromagnet with two sublattice spins aligned along the $z$-direction, $\mathbf{S}_s$ with $s\in\{1,2\}$, the Hamiltonian can be written as

\begin{equation}\label{eq:spinhamiltonian}
    H_0 = J\mathbf{S}_1\cdot\mathbf{S}_2 + \sum_{s=1,2} ( D_x S_{s,x}^2 + D_y S_{s,y}^2),
\end{equation}
where $J$ is the antiferromagnetic exchange interaction and $D_x$ and $D_y$ are the anisotropy energies. For $D_x < D_y$, this system hosts low- and high-frequency magnon modes with eigenfrequencies of $\hbar\Omega_{l} = 2\sqrt{(J+D_y)D_x}$ and $\hbar\Omega_{h} = 2\sqrt{(J+D_x)D_y}$. The spin configuration is shown in Fig.~\ref{fig:setup}. 

At the same time, the full phonon-dependent potential is given by

\begin{equation}
    V_{ph} = V_0 + V_{s\text{-}ph} + V_{l\text{-}m}.
\end{equation}
The first term, $V_0 = \Omega_0^2 ( Q_y^2 + Q_z^2)/2$, is the potential energy of a doubly degenerate phonon mode in the $yz$-plane, where $\Omega_0$ is the eigenfrequency, and $Q_y$ and $Q_z$ are the phonon amplitudes to the two orthogonal components. A circular superposition of these components results in a chiral phonon mode, as shown in Fig.~\ref{fig:setup}. Note that these degenerate chiral phonon modes live at the center of the Brillouin zone, where they can be excited with light. Nondegenerate chiral phonon modes, such as in hexagonal 2D materials or chiral crystals \cite{zhang:2015,Zhu2018,Ishito2023,Ueda2023}, are not considered here. 

The second term, $V_{s\text{-}ph}=-\mathbf{m}\cdot\mathbf{B}_{ph}$, describes the interaction between the spin and lattice degrees of freedom, which is given by the coupling of the effective magnetic field produced by the chiral phonon mode, $\mathbf{B}_{ph}$, to the magnetic moment of the magnon mode, $\mathbf{m}$, which can be considered a phonon inverse Faraday or phonon Barnett effect \cite{Juraschek2020_3,Davies2023}.
$\mathbf{B}_{ph} = \mu_0 \gamma_{ph} \mathbf{L}_{ph}/V_c$, where $\gamma_{ph}$ is the gyromagnetic ratio of the chiral phonon, $\mu_0$ is the vacuum permeability, and $V_c$ is the unit-cell volume. $\mathbf{L}_{ph}=\mathbf{Q}\times\dot{\mathbf{Q}}$ is the phonon angular momentum, where $\mathbf{Q}=(0,Q_y,Q_z)$. The magnetic moment of the magnon mode is given by $\mathbf{m}=\gamma_{el} \hbar(\mathbf{S}_1+\mathbf{S}_2)$, where $\gamma_{el}$ is the gyromagnetic ratio of the electron. Using these relations, the potential can also be written as an angular momentum coupling, $V_{s\text{-}ph}=k \mathbf{L}_{ph}\cdot\mathbf{S}$, where $k=-\mu_0\gamma_{ph}\gamma_{el}\hbar/V_c$ is the coupling coefficient. For the purpose of our analysis and without loss of generality, we choose the chiral phonon mode to be circularly polarized in the $yz$-plane ($\mathbf{L}_{ph}||\hat{\mathbf{x}}$), which allows it to couple to the magnetic moment of the high-frequency magnon mode, $\mathbf{m}=(m_x,0,0)$, as illustrated in Fig.~\ref{fig:setup}. For a coupling to the low-frequency magnon mode, the chiral phonon mode would need to be circularly polarized in the $xz$-plane ($\mathbf{L}_{ph}||\hat{\mathbf{y}}$). 

The third term, $V_{l\text{-}m} = -(\mathbf{p}_y + \mathbf{p}_z)\cdot\mathbf{E}(t)$, describes the light-matter interaction of the chiral phonon mode with the laser pulse and is given by electric-dipole coupling. $\mathbf{p}_n=\mathbf{Z}_n Q_n$ is the electric dipole moment of the phonon component $n\in\{y,z\}$, $\mathbf{Z}_n$ is the mode effective charge, and $\mathbf{E}(t)$ is the electric field component of the laser pulse. We will use the Einstein sum convention for summing over indices that denote spatial directions throughout the manuscript. The components $Z_{n,i}$ are nonzero if the driven phonon mode is infrared active and the coordinate system can be chosen such that $n\equiv i$.

The coherent spin dynamics are described by the Landau-Lifshitz-Gilbert equation \cite{Kampfrath2011,Fechner2018,Rezende2019,Juraschek2021}, 

\begin{equation} \label{eq:LLG}
    \frac{{\rm d}\mathbf{S}_s}{{\rm d}t}=\frac{\gamma_{el}}{1+\kappa_{el}^2}\Big[\mathbf{S}_s \times \mathbf{B}_s^{\rm eff} - \frac{\kappa_{el}}{|\mathbf{S}_s|} \mathbf{S}_s \times (\mathbf{S}_s \times \mathbf{B}_s^{\rm eff})\Big],
\end{equation}
where $\kappa_{el}$ is the Gilbert damping and the spin vectors are normalized, $|\mathbf{S}_s|=1$. $\mathbf{B}^\mathrm{eff}_s = -(\gamma_{el}\hbar)^{-1}\partial_{\mathbf{S}_s} H$ is the effective magnetic field acting on the spins of sublattice $s$, with the full spin-dependent Hamiltonian given by $H=H_0+V_{s\text{-}ph}$. The effective magnetic field is given by $\mathbf{B}_s^\mathrm{eff} = \mathbf{B}_{ph}-(\gamma_{el}\hbar)^{-1}(J\mathbf{S}_{s'}+2D_x S_{s,x}\hat{\mathbf{x}}+2D_y S_{s,y}\hat{\mathbf{y}})$, where $s,s'\in\{1,2\}$ and $s\neq s'$. The coherently driven chiral phonon mode therefore enters the Landau-Lifshitz-Gilbert equations as a driving force given by $\mathbf{B}_{ph}$. 

The coherent chiral phonon dynamics can in turn be described by a phenomenological oscillator model \cite{Fechner2018,Juraschek2021}, $\ddot{Q}_n + \kappa_n \dot{Q}_n + \partial_{Q_n}V_{ph} = 0$, where $\kappa_y=\kappa_z\equiv\kappa_0$ is the phonon linewidth. The equations of motion yield

\begin{align}
    \ddot{Q}_n & + \kappa_0 \dot{Q}_n + \Omega_0^2{Q}_n = Z_{n,i}  E_i(t) \nonumber\\
    & \pm k\left[ Q_{n'}(\dot{S}_{1,x}+\dot{S}_{2,x})+2\dot{Q}_{n'}(S_{1,x}+S_{2,x}) \right], \label{eq:backaction}
\end{align}
where $n\neq n'$ and the second line of Eq.~\eqref{eq:backaction} describes the back-action of the spin system on the phonon dynamics.

\subsection{Quasistatic magnetization and nonlinear magnetoelectric susceptibility}

In the case of nonlinear phononic rectification, the minimum of the coupled phonon amplitude is shifted away from its equilibrium position, $Q_{c,\mathrm{min}}\neq 0$ (see Appendix~A). We now derive an analog expression for the minimum of the magnetization of the magnon mode that obtains a nonzero value, $M_{x,\mathrm{min}}\neq 0$. 

We minimize the full spin-dependent Hamiltonian, $H=H_0+V_{s\text{-}ph}$, under the constraint of normalized spin vectors using Lagrange multipliers, $\partial_{\mathbf{S}_s} [H + \lambda_1(|\mathbf{S}_1|-1) +\lambda_2(|\mathbf{S}_2|-1)] = 0$. (Please see Appendix~B for details of the derivation.) Using $M_x=m_x/V_c$ for the magnetization, the calculation yields

\begin{equation}
    M_{x,\mathrm{min}} = -\frac{\gamma_{el}\hbar}{V_c}\frac{k}{J+D_x}L_{ph,x}, \label{eq:minimummag}
\end{equation}
which means that the spin precession is rectified such that the average value of the induced magnetization is proportional to the phonon angular momentum, $\langle M_x\rangle \propto \langle L_{ph,x} \rangle$. This corresponds to a quasistatic canting of the spins in the presence of the phonon angular momentum that can be described by an induced mean tilt angle, $\langle \theta \rangle \neq 0$. The relation is valid for the regime of small-angle deviations ($\theta\ll 90^\circ$), and breaks down for large $L_{ph,x}$. Eq.~\eqref{eq:minimummag} is the first part of the main result of our study.

In order to obtain the full spin-dynamical response to the coherent excitation of the chiral phonon modes, we solve the Landau-Lifshitz-Gilbert equations in frequency space to linear order in the effective phono-magnetic field. (See Appendix B for details.) Because each of the orthogonal components of the chiral phonon mode is driven linearly by the electric field component of the laser pulse in Eq.~\eqref{eq:backaction}, the effective phono-magnetic field can be expressed to quadratic order in the electric field as

\begin{equation}
    B_{ph,x}(\omega) = \frac{\mu_0\gamma_{ph}Z_jZ_k}{V_c}\epsilon_{xjk}\frac{E_j(\omega)}{\Delta_0(\omega)} \circledast \frac{i\omega E_k(\omega)}{\Delta_0(\omega)},    \label{eq:magneticfieldsolution}
\end{equation}
where $\Delta_0(\omega)=\Omega_0^2-\omega^2+i\kappa_0\omega$, and $\circledast$ is the convolution operator. With the effective phono-magnetic field acting as a driving force for the Landau-Lifshitz-Gilbert equations \eqref{eq:LLG}, we can derive analytical approximations for the spin components $S_{1x}=S_{2x}=S_x$ and $S_{1y}=-S_{2y}=S_y$ in linear order of $B_{ph,x}$. This results in a quadratic dependence of the spin dynamics on the electric field through Eq.~\eqref{eq:magneticfieldsolution}. We obtain the solutions
\begin{align}
S_x(\omega) &= \frac{\gamma_{el}}{1+\kappa_{el}^2}\frac{i\kappa_{el}\omega+2 D_y/\hbar}{\Delta_m(\omega)}B_{ph,x}(\omega), \label{eq:Ssol1} \\
S_y(\omega) &= \pm \frac{\kappa_{el}\gamma_{el}}{1+\kappa_{el}^2}\frac{i\omega}{\Delta_m(\omega)}B_{ph,x}(\omega),\label{eq:Ssol2}
\end{align}
where we have defined $\Delta_m(\omega)=\Omega_m^2-\omega^2+i\kappa_m\omega$. $\Omega_m=\frac{2}{\hbar}\sqrt{\frac{(J+D_x)D_y}{1+\kappa_{el}^2}}$ is the damping-renormalized magnon frequency and $\kappa_m=\frac{2\kappa_{el}}{\hbar\left(1+\kappa_{el}^2\right)}\left(J+D_x+D_y\right)$ is the magnon linewidth. 

Using $M_x(\omega) = 2\gamma_{el}\hbar S_x(\omega)/V_c$ in combination with Eq.~\eqref{eq:magneticfieldsolution} (see Appendix B for details), we obtain

\begin{equation}\label{eq:magnetoelectricintegral}
    M_x(\omega) = \frac{1}{\sqrt{2\pi}} \int_{-\infty}^{\infty}\chi^{(2)}_{me,xjk}(\omega,\omega')E_j(\omega-\omega')E_k(\omega')d\omega',
\end{equation}
where $\chi^{(2)}_{me,xjk}$ is the second-order nonlinear magnetoelectric susceptibility, given by

\begin{align}
\chi^{(2)}_{me,xjk}(\omega,\omega') = & \frac{2\mu_0\gamma_{ph}\gamma_{el}^2\hbar\epsilon_{xjk} Z_yZ_k}{V_c^2\left(1+\kappa_{el}^2\right)} \nonumber\\
& \times \frac{\left(i\kappa_{el}\omega+2 D_y/\hbar\right)i\omega'}{\Delta_m(\omega)\Delta_0(\omega-\omega')\Delta_0(\omega')}. \label{eq:magnetoelectricsusceptibility}
\end{align}
For a resonantly driven chiral phonon mode, $\omega'=\Omega_0$, and the nonlinear magnetoelectric susceptibility reduces to $\chi^{(2)}_{me,xyz}(\omega,\Omega_0)$. Eq.~\eqref{eq:magnetoelectricsusceptibility} is the second part of the main result of our study.

\begin{figure*}[t]
    \centering
    \includegraphics[scale=0.125]{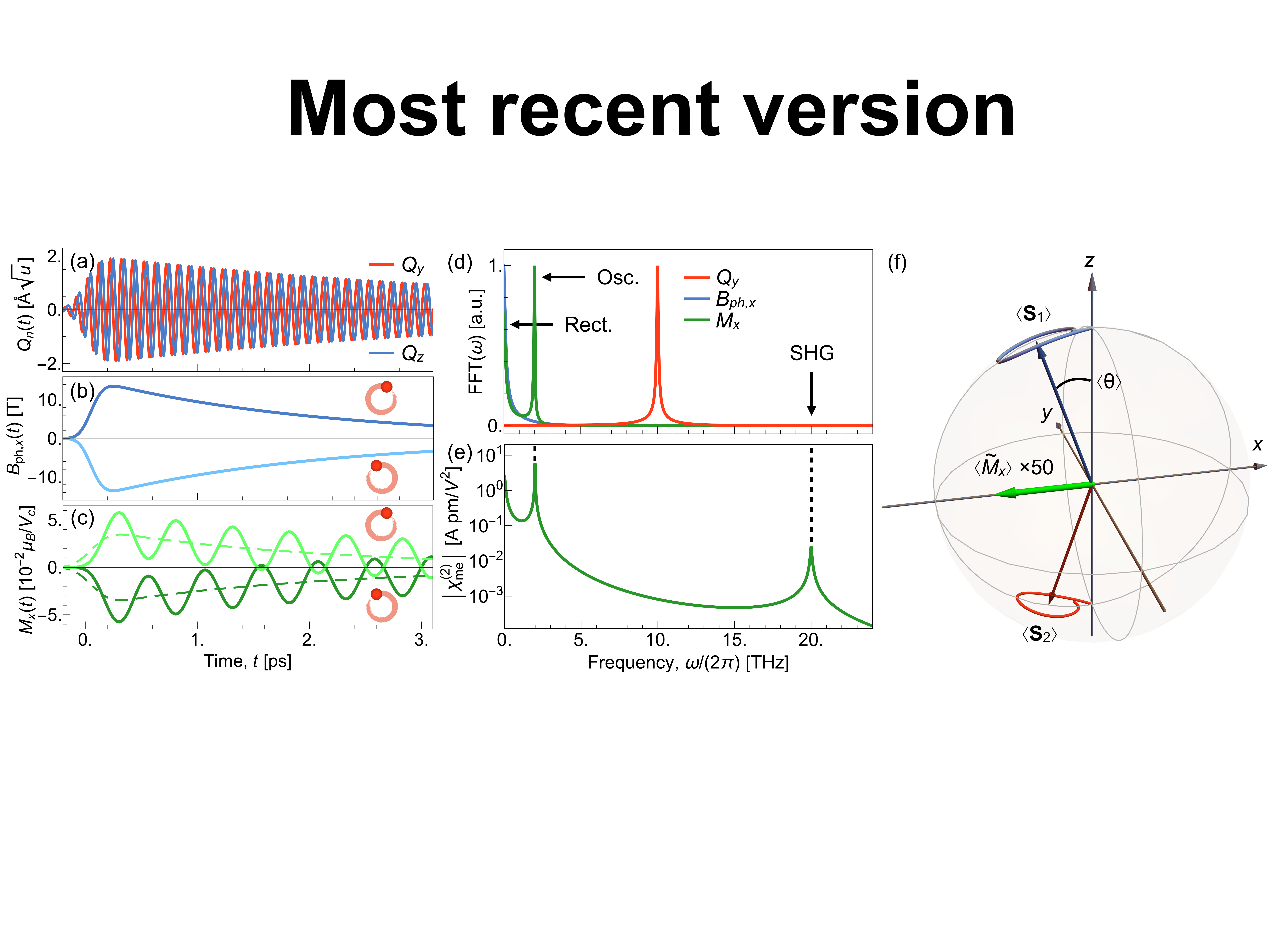}
    \caption{
    Nonlinear magnonic rectification. (a) Time-dependent phonon amplitudes of the two components of the chiral phonon mode, $Q_y$ and $Q_z$, excited by a circularly polarized laser pulse at $t=0$. We use a central frequency of $\omega_0=10$~THz, a peak electric field of $E_0/\sqrt{2}=10$~MV/cm, and a FWHM duration of $\tau=250$~fs. (b) Time evolution of the effective magnetic field generated by the chiral phonon mode, $B_{ph,x}$. The direction of the phono-magnetic field depends on the helicity of the chiral phonon mode. (c) Time-dependent magnetization, $M_x$, induced by the spin precession of the high-frequency magnon mode, dependent on the helicity of the excitation. The dashed line shows the quasistatic DC component of the magnetization. (d) Normalized Fourier transforms of the time traces in (a-c). Shown are the chiral-phonon resonance at 10~THz and the static component of the effective phono-magnetic field at zero frequency. The spectrum of the magnon mode shows both a resonance at its eigenfrequency of 2~THz, corresponding to the oscillatory part, as well as a rectified, zero-frequency component, corresponding to the quasistatic magnetization. (e) Absolute value of the nonlinear magnetoelectric susceptibility under resonant driving conditions of the chiral phonon mode, $|\chi_{me,xyz}^{(2)}(\omega,\Omega_0)|$. In addition to the rectified and oscillatory components, the magnon exhibits a high-frequency response at the double frequency of the chiral phonon mode, which corresponds to second-harmonic generation. This component is orders of magnitude smaller than the others and therefore not visible in the linear scaling of the Fourier spectrum in (d). (f) Spin precession along the eigenvectors of the high-frequency magnon mode. Red and blue arrows show the transient tilting of the axis of spin precession by an angle $\langle\theta\rangle$. The trajectories on the Bloch sphere cover the first full precession period. The green arrow shows the quasistatic component of magnetization, $\langle \tilde{M}_x\rangle = \langle M_x\rangle/(\gamma_{el}\hbar)$, exaggerated by a factor of 50.
    }
    \label{fig:spindynamics}
\end{figure*}

\subsection{Simulation of light-induced weak ferromagnetism}

We now evaluate the coupled spin-lattice dynamics that follow the excitation of the chiral phonon mode by an ultrashort laser pulse, as described by Eqs.~\eqref{eq:LLG} and \eqref{eq:backaction} and following the setup in Fig.~\ref{fig:setup}. We use common values for the parameters of the spin Hamiltonian and for the chiral phonon mode \cite{Kampfrath2011,Rezende2019,Juraschek2019,Geilhufe2021}. We set $J=106$~meV, $D_y=160$~$\mu$eV, and $D_x=3.5$~$\mu$eV to model a system with a magnon eigenfrequency of $\Omega_h/(2\pi)=2$~THz. The Gilbert damping is taken to be $\kappa_{el}=2.4\cdot 10^{-4}$ and the electronic gyromagnetic ratio of the magnetic ion $\gamma_{el}=-2.2\cdot 10^{11}$~(Ts)$^{-1}$. We use an eigenfrequency of $\Omega_0/(2\pi)=10$~THz for the chiral phonon mode and set the mode effective charges to $Z_{n,i}=1~e/\sqrt{u}$, as well as the phenomenological damping to $\kappa_0=0.05\Omega_0/(2\pi)$. We set the unit-cell volume to $V_c=35$~\AA$^3$. The only remaining parameter is the phonon gyromagnetic ratio, for which experimental and theoretical values have only been described recently. Initial calculations were suggesting values of $\gamma_{ph}\sim 10^{7}$~(Ts)$^{-1}$, corresponding to a phonon magnetic moment of $\mu_{ph}=\gamma_{ph}\hbar\sim 10^{-4}\mu_B$ \cite{juraschek2:2017,Juraschek2019,Juraschek2020_3,Geilhufe2021,Zabalo2022}. Recent experiments and theoretical predictions have however shown that, depending on the microscopic origin of the spin-phonon interaction \cite{Cheng2020,Ren2021,Juraschek2022_giantphonomag,Basini2022,Davies2023,Baydin2022,hernandez2022chiral,Geilhufe2023,Chaudhary2023,Wu2023_phononmagneticmoments}, phonon magnetic moments of up to several $\mu_B$ exist that correspond to a phonon gyromagnetic ratio of $\gamma_{ph}\sim 10^{11}$ to $10^{12}$~(Ts)$^{-1}$, and we use a moderate value of $10^{11}$~(Ts)$^{-1}$ in our calculations. The circularly polarized electric field component of the ultrashort laser pulse is given by $\mathbf{E}(t)=(E(t)\cos(\omega_0 t),E(t)\sin(\omega_0 t),0)/\sqrt{2}$, with $E(t) = E_0 \exp[-t^2/(2(\tau_0/\sqrt{8\ln2})^2)]$. We use a central frequency of $\omega_0/(2\pi)=10$~THz, a peak electric field of $E_0/\sqrt{2}=10$~MV/cm, and a FWHM duration of $\tau=250$~fs, parameters which lie well within experimentally achievable pulse intensities \cite{Liu2020,Vicario2020}.

In Fig.~\ref{fig:spindynamics}, we present the results of the dynamical simulations. Fig.~\ref{fig:spindynamics}(a) shows the time evolution of the phonon amplitudes, $Q_y$ and $Q_z$, that oscillate with a $\phi=\pi/2$ phase shift, resulting in circular polarization. In Fig.~\ref{fig:spindynamics}(b), we show the time evolution of the effective magnetic field generated by the chiral phonon mode, $B_{ph,x}$, which reaches a maximum amplitude of 13.5~T before decaying quadratically as a function of the phonon amplitudes. The direction of the magnetic field can be reversed by changing the helicity of the laser pulse and therefore of the chiral phonon mode. In Fig.~\ref{fig:spindynamics}(c), we show the time evolution of the magnetization that is induced by the spin precession of the high-frequency magnon mode for both left- and right-handed circular excitations. The quasistatic shift away from the equilibrium position is clearly visible, reaching a transient maximum of $\langle M_x\rangle=0.03$~$\mu_B$/$V_c$. The quasistatic component decays on the timescale of the effective phono-magnetic field. In Fig.~\ref{fig:spindynamics}(d), we show the normalized Fourier transforms of the time traces in (a-c). There is a sharp and symmetric peak at the eigenfrequency of the chiral phonon mode, which points towards the fact that the back-action of the spin system in the phonon equations of motion~\eqref{eq:backaction} is negligible. The effective phono-magnetic field shows a broadened static component at zero frequency, corresponding to the unidirectional force it exerts on the spins. The Fourier transform of the magnon mode shows three contributions: first, a sharp peak at its eigenfrequency, corresponding to the oscillatory spin precession, second, a static component at zero frequency, corresponding to the rectified, quasistatic shift, and third, a component at the double frequency of the chiral phonon mode, $2\Omega_0$, corresponding to second harmonic generation. The latter feature is not resolved in the linear scaling of the Fourier spectrum, but can be seen in Fig.~\ref{fig:spindynamics}(e), where we show the absolute value of the second-order nonlinear magnetoelectric susceptibility from Eq.~\eqref{eq:magnetoelectricsusceptibility}, $|\chi^{(2)}_{me,xyz}(\omega,\Omega_0)|$, on a logrithmic scale. The three features of rectification, oscillation, and second harmonic generation are clearly visible. Finally, in Fig.~\ref{fig:spindynamics}(f), we show the trajectories of the sublattice spin precessions, $\mathbf{S}_1$ and $\mathbf{S}_2$, for the time period of the first full precession. The spin arrows indicate the quasistatic spin canting and tilt of the axis of spin precession, amounting to an angle of $\langle\theta\rangle=0.5^\circ$ at its maximum. Again, the direction of spin canting and therefore of the tilt angle can be reversed by changing the helicity of the chiral-phonon excitation. This is the third and final part of the main result of our study.


\section{Discussion}

We have shown that a transient spin canting and therefore a quasistatic magnetization can be induced in an antiferromagnet through the coherent excitation of a chiral phonon mode, a mechanism we call nonlinear magnonic rectification. We predict that the axis of spin precession can be transiently tilted by an angle on the order of $\langle\theta\rangle=0.5^\circ$, which is comparable to the spin canting commonly found in weak ferromagnets \cite{Eibschutz1967}. The rectification mechanism therefore leads to a realization of light-induced transient weak ferromagnetism. While we have considered the example of a Heisenberg antiferromagnet with easy-plane anisotropy in our study, our results are readily extendable and general to all magnetic materials that possess degenerate chiral phonon modes with the required symmetry to couple to the corresponding magnon mode. This applies to a wide variety of materials within the cubic, tetragonal, hexagonal, and trigonal crystal systems. The mechanism is further applicable to elliptically polarized nearly-degenerate phonon modes in crystal systems without in-plane symmetry (orthorhombic, monoclinic, and triclinic), such as NiO \cite{Tzschaschel2017}, as long as the dephasing time of the two components is longer than the phonon lifetime, $2\pi(\Omega_y-\Omega_z)^{-1} > \kappa_{y/z}^{-1}$.

Within recent years, the various microscopic origins that underly the angular momentum coupling between chiral phonons and electron spins have been under heavy investigation \cite{nova:2017,juraschek2:2017,Shin2018,Cheng2020,Hamada2020,Juraschek2020_3,Geilhufe2021,Streib2021,Ren2021,Juraschek2022_giantphonomag,Basini2022,Baydin2022,hernandez2022chiral,Fransson2022,Geilhufe2023,Davies2023,Zhang2023chiral}, and the list is rapidly expanding. Our results are agnostic towards the microscopic mechanism, as long as it is compatible with the macroscopic symmetry requirement presented by the spin-chiral phonon coupling, $V_{s\text{-}ph}$. A particularly interesting candidate is CoTiO$_3$, an antiferromagnetic transition-metal oxide, which hosts infrared-active degnerate chiral phonon modes that has very recently been predicted to possess large phonon magnetic moments and therefore strong coupling to spin degrees of freedom \cite{Lujan2023,Chaudhary2023}. As circularly polarized terahertz pulses with high electric field strenghts have become feasible lately \cite{Basini2022,Davies2023}, we expect that nonlinear magnonic rectification can be realized with state-of-the-art tabletop pump-probe setups.

We distinguish the mechanism from other forms of spin excitation and control. A direct excitation of the magnon mode through Zeeman coupling to the magnetic field component of light can be excluded \cite{Kampfrath2011,Baierl2016_2}, because our simulated ultrashort pulses are far off resonance from the magnon eigenfrequencies. Furthermore, the laser pulse can excite the magnon modes through the inverse Faraday or Cotton-Mouton effects \cite{Kalashnikova2007,Tzschaschel2017}, which will always happen in parallel to the proposed mechanism here, if the selection rules of the magnetic Raman tensor allow for it. Generally, opto-magnetic effects produce effective magnetic fields that may induce low-frequency spin dynamics \cite{Blank2022_spin_precession,Blank2023}. While this contributes to the amplitude of spin precession, it will not induce a rectification of the axis of precession. A third type of mechanism related to the context is the transient generation of magnetization through piezomagnetically active phonon modes \cite{Radaelli2018,Disa2020}. Here, the magnetization is induced through a rectified phonon mode however, with no magnon excitation and spin precession involved, which falls under nonlinear phononic rectification. Finally, there has been a report of nonlinear magnon dynamics by direct excitation of spin precession through the magnetic field component of light, however this mechanism does not go along with rectification \cite{Mashkovich2021,Kurihara2022,Zhang2022_nonlmag,Zhang2023_nonlmag}.

To conclude, the mechanism of nonlinear magnonic rectification proposed here enables the possibility to generate a quasistatic magnetization from spin precession and represents a third fundamental rectification process besides optical and nonlinear phononic rectification. Our example of light-induced weak ferromagnetism can be seen in the same light as the light-induced generation and control of magnetic order mediated by chiral phonons, demonstrated in recent experiments \cite{Luo2023,Basini2022,Davies2023}.


\begin{acknowledgments}
T.K. and D.A.B.L. contributed equally to this work. We are grateful to Wanzheng Hu for useful discussions. T.K. and D.M.J. acknowledge support from Tel Aviv University. D.A.B.L. acknowledges support from
the U.S. Department of Energy, Office of Science, Office of Basic Energy Sciences Early Career Research Program under Award Number DE-SC-0021305. Calculations were performed on local computing infrastructure at Tel Aviv University.
\end{acknowledgments}


\section*{Appendix A: Nonlinear phononic rectification at the example of bismuth ferrite}

\subsection*{Quasistatic distortion and nonlinear electric susceptibility}

Here, we present an example of nonlinear phononic rectification. We will review the general theory and then show that the rectification can be utilized to induce a transient ferroelectric polarization in noncentrosymmetric materials. A minimal model consists of two phonon modes, one driven ($d$) and one that is coupled to it nonlinearly ($c$) through three-phonon coupling \cite{subedi:2014,Juraschek2018},

\begin{equation}\label{eq:minimalpotential}
V_{ph} = \frac{\Omega^2_d}{2}Q_d^2 + \frac{\Omega^2_c}{2}Q_c^2 + c Q_d^2 Q_c.
\end{equation}

Here, $\Omega_{d/c}$ are the eigenfrequencies of the driven and coupled phonon modes and $c$ is the nonlinear coupling coefficient. The presence of the coupling term modifies the energy minimum of the coupled mode, which we obtain from $\partial_{Q_c} V_{ph}=0$ as 

\begin{equation}\label{eq:phononminimumAPPENDIX}
    Q_{c,\mathrm{min}} = -cQ_d^2/\Omega_c^2.
\end{equation}

The light-matter interaction is given by electric-dipole coupling,

\begin{equation}\label{eq:light-matter}
V_{l\text{-}m} = -\mathbf{p}_d\cdot\mathbf{E}(t) = - Z_{d,i} Q_d E_i(t),
\end{equation}
where $\mathbf{p}_d=\mathbf{Z}_d Q_d$ is the electric dipole moment of the driven phonon mode, $\mathbf{Z}_d$ is the mode effective charge, and $\mathbf{E}(t)$ is the electric field component of the laser pulse. We will use the Einstein sum convention for summing over indices that denote spatial directions. We assume that the polarization of the laser pulse and the phonon dipole moment are aligned in parallel. The components $Z_{d,i}$ are nonzero if the driven phonon mode is infrared active.

The coherent dynamics of the coupled phonon modes can be described by a phenomenological oscillator model \cite{subedi:2014,subedi:2015,fechner:2016,juraschek2:2017,Juraschek2018}, $\ddot{Q}_n + \kappa_n \dot{Q}_n + \partial_{Q_n}V = 0$, where $\kappa_n$ is the linewidth of the phonon mode, $n\in\{d,c\}$, and $V=V_{ph}+V_{l\text{-}m}$ is the total potential energy. This results in two coupled equations of motion, 

\begin{align}
    \ddot{Q}_d + \kappa_d \dot{Q}_d + (\Omega_d^2 + 2cQ_c) Q_d & = Z_{d,i} E_i(t), \label{eq:drivenmode}\\
    \ddot{Q}_c + \kappa_c \dot{Q}_c + \Omega_c^2 Q_c & = -c Q_d^2. \label{eq:coupledmode}
\end{align}

While the electric field component of light, $E_i$, serves as the driving force for the driven mode, $Q_d^2$ serves as driving force for the coupled mode. In frequency space, we can obtain analytical expressions for $Q_d$ and $Q_c$ in first and second order of the electric field, respectively, which read

\begin{align}
   Q_d(\omega) & = Z_{d,i} \frac{E_{i}(\omega)}{\Delta_d(\omega)}, \label{eq:phononsolution1}\\
   Q_c(\omega) & =-\frac{c }{\Delta_c(\omega)} {Q}_d(\omega)\circledast  {Q}_d(\omega) \nonumber\\ &= -\frac{cZ_{d,i}Z_{d,j} }{\Delta_c(\omega)} \Bigg(\frac{{E}_i(\omega)}{\Delta_d(\omega)}\circledast \frac{{E}_j(\omega)}{\Delta_d(\omega)}\Bigg),\label{eq:phononsolution2}
\end{align}
where $\Delta_{d/c}(\omega)=\Omega_{d/c}^2 - \omega^2 + i\omega\kappa_{d/c}$ and $\circledast$ is the convolution operator. 

In noncentrosymmetric materials, the coupled phonon mode can also be infrared active, $\mathbf{p}_c=\mathbf{Z}_c Q_c$. The electric polarization induced by the coupled phonon mode, $P_{c,i}=p_{c,i}/V_c$, where $V_c$ is the unit-cell volume of the crystal, can then be expressed as

\begin{align}\label{eq:second-ordersusceptibility}
P_{c,i}(\omega)  = \dfrac{\varepsilon_0}{\sqrt{2\pi}}\int_{-\infty}^{\infty}\chi_{e,ijk}^{(2)}(\omega,\omega'){E}_j(\omega-\omega'){E}_k(\omega')d\omega',
\end{align}
where $\chi_{e,ijk}^{(2)}$ is the second-order nonlinear electric susceptibility induced by the nonlinear phonon coupling, given by
\begin{align}
    \chi_{e,ijk}^{(2)}(\omega,\omega')=-\frac{c }{\varepsilon_0V_c}\frac{Z_{c,i} Z_{d,j} Z_{d,k}}{\Delta_c(\omega)\Delta_d(\omega-\omega')\Delta_d(\omega')}.
\end{align}
Note that we neglect the direct contribution for the coupled phonon mode as in Eq.~\eqref{eq:phononsolution1}, because we assume the laser pulse to be tuned into resonance with the driven mode, with no significant spectral overlap with the coupled mode.

\subsection*{Simulation of nonlinear phonon dynamics}

We now evaluate the mechanism for the concrete example of BiFeO$_3$, a perovskite oxide (point group $3m$) that exhibits a ferroelectric polarization along the direction of rhombohedral distortion at room temperature \cite{Neaton_et_al:2005}. BiFeO$_3$ hosts fully symmetric and infrared-active phonon modes with the irreducible representation $A_1$, whose electric dipole moments are aligned parallel to the ferroelectric polarization, and which are allowed to couple according to $Q_d^2 Q_c$, where $d$ and $c$ correspond to different $A_1$ modes in the system. We pick the illustrative example of the coupling of the highest-frequency $A_1$ mode ($d$) at 15.3~THz and the lowest-frequency $A_1$ mode ($c$) at 4.8~THz in the system. We use the parameters previously calculated in Ref.~\cite{BustamanteLopez2023} with a nonlinear coupling coefficient of $c=-34$~meV/(\AA $\sqrt{u}$)$^3$, where $u$ is the atomic mass unit, and mode effective charges of $Z_{d,z}=-0.55~e/\sqrt{u}$ and $Z_{c,z} = 0.69~e/\sqrt{u}$, where $e$ is the elementary charge. For the linewidths, we assume phenomenological values for transition-metal oxides of $\kappa_n = 0.1\Omega_n/(2\pi)$ \cite{fechner:2016,juraschek2:2017}. We model the laser pulse as $E_z(t) = E_0 \exp[-t^2/(2(\tau_0/\sqrt{8\ln2})^2)] \cos(\omega_0 t)$, where $E_0=15$~MV/cm is the peak electric field, $\tau_0=250$~fs is the FWHM duration, and $\omega_0/(2\pi)=15.3$~THz is the center frequency of the laser pulse. 

\begin{figure}[t]
    \centering
    \includegraphics[scale=0.091]{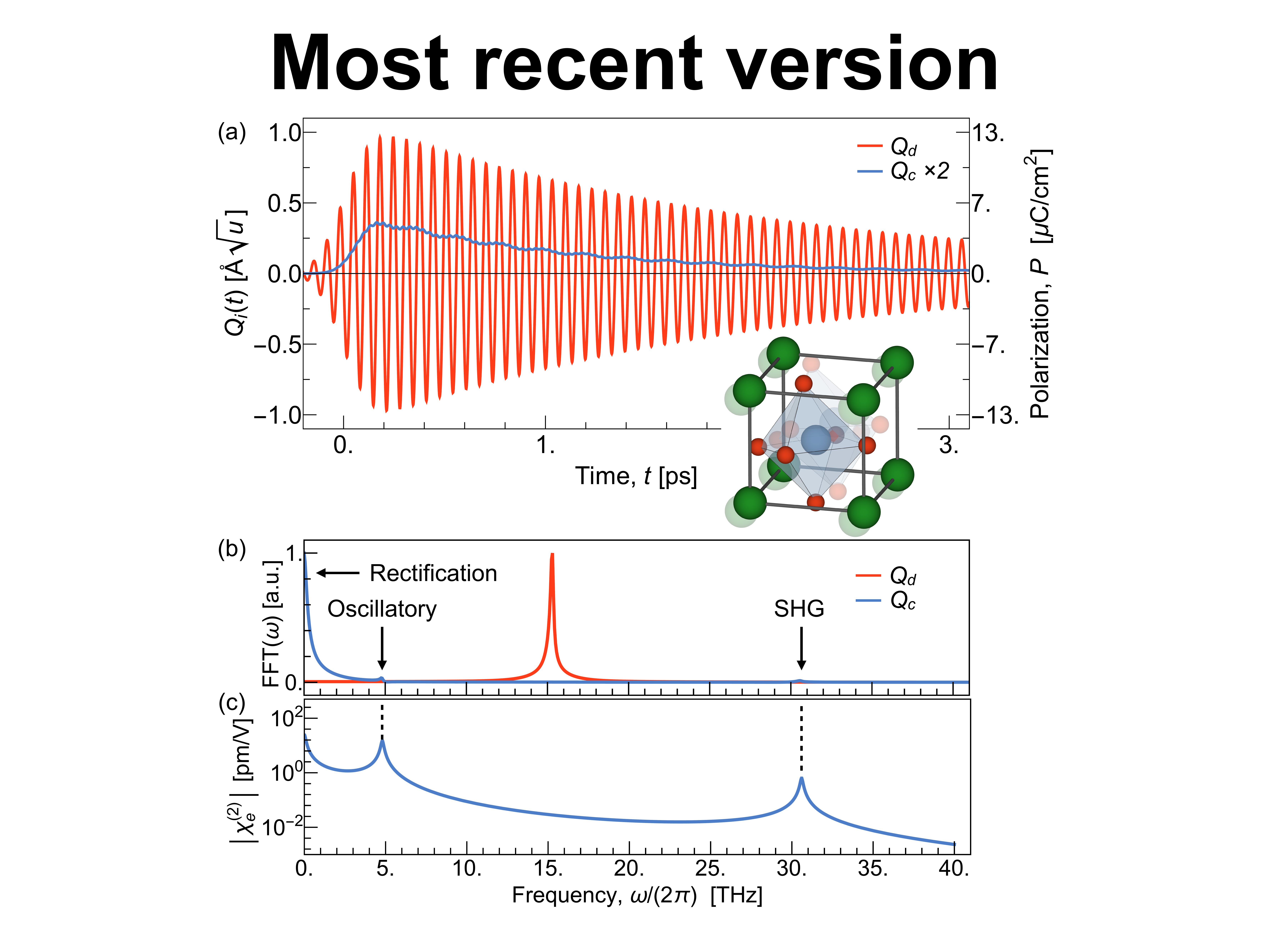}
    \caption{Nonlinear phononic rectification. (a) Time evolution of the phonon amplitudes of the driven $A_1(15.3)$ mode ($Q_d$) and the coupled $A_1(4.8)$ mode ($Q_c$) in BiFeO$_3$, displaying the quasistatic shift of the coupled phonon mode and the corresponding induced ferroelectric polarization, $P_{c,z}$. The ultrashort pulse is centered at $t=0$, with a central frequency of $15.3$~THz, peak electric field of $15$~MV/cm, and a FWHM duration of $250$~fs. Inset shows the quasistatic distortion caused by the $A_1(4.8)$ mode (exaggerated). (b) Normalized Fourier transform of the phonon amplitudes, showing the resonance of the $A_1$(15.3) mode and the static component of the $A_1$(4.8) mode at zero frequency. (c) Real and imaginary parts of the second-order electric susceptibility, $\chi^{(2)}_e$, arising from the nonlinear phonon coupling. Displayed are the quadratic and simple poles at the eigenfrequencies of the driven $A_1$(15.3) and coupled $A_1$(4.8) mode, respectively, as described by Eq.~\eqref{eq:second-ordersusceptibility}.}
    \label{fig:phonondynamics}
\end{figure}

The results of the phonon dynamics are shown in Fig.~\ref{fig:phonondynamics}. Fig.~\ref{fig:phonondynamics}(a) shows the time evolutions of the phonon amplitudes, $Q_d$ and $Q_c$, in response to the excitation by the laser pulse. The rectification of the coupled phonon mode is clearly visible. In Fig.~\ref{fig:phonondynamics}(b), we show the normalized Fourier spectra of the time traces in (a), which reveal a clear resonance of the driven phonon mode at its eigenfrequency and a clear peak of the coupled phonon mode at zero frequency, corresponding to nonlinear phononic rectification (quasistatic distortion). The coupled-mode spectrum reveals two additional, but much smaller peaks, one at its eigenfrequency, which corresponds to the oscillatory response of the coupled mode, and one at the double frequency of the driven mode, which corresponds to second harmonic generation. These responses become more pronounced in a logarithmic plot of the second-order nonlinear electric susceptibility, $\chi_{e}^{(2)}$, which is shown in Fig.~\ref{fig:phonondynamics}(c). $\chi_{e}^{(2)}$ clearly reveals the three distinct responses arising from nonlinear coupling of $Q_c$ and $Q_d$.


\section*{Appendix B: Nonlinear magnonic rectification - derivations}

We derive an expression for the quasistatic shift of the spin-precession axis that results in a transient magnetization, as well as for the second-order nonlinear magnetoelectric susceptibility describing rectification, oscillatory response, and second harmonic generation.

\subsection*{Model of spin-chiral phonon coupling}

The coupled spin and lattice degrees of freedom can be described by the Hamiltonian $H=H_0 + V_{ph} + V_{s\text{-}ph} + V_{l\text{-}m}$. Here, $H_0$ is the ground-state spin Hamiltonian for a Heisenberg antiferromagnet with easy-plane anisotropy, $V_{ph}$ is the potential energy of the doubly degenerate phonon mode, $V_{s\text{-}ph}$ is the spin-chiral phonon coupling, and $V_{l\text{-}m}$ is the electric-dipole coupling to the electric field component of light. We follow the setup displayed in Fig.~\ref{fig:setup} in the main text. The four contributions are given by

\begin{align}
    H_0 & = J\mathbf{S}_1\cdot\mathbf{S}_2 + \sum_{s=1,2} (D_x S_{s,x}^2 + D_y S_{s,y}^2), \label{eq:spinhamiltonianAPPENDIX} \\
    V_{ph} & = \frac{\Omega_0^2}{2} \left( Q_y^2 + Q_z^2 \right) \label{eq:chiralphononpotentialAPPENDIX} \\
    V_{s\text{-}ph} & = -\mathbf{m}\cdot\mathbf{B}_{ph}, \label{eq:spin-phononAPPENDIX} \\
    V_{l\text{-}m} & = -(\mathbf{p}_y+\mathbf{p}_z)\cdot\mathbf{E}(t). \label{eq:light-matterAPPENDIX}
\end{align}

Here, $J$ is the exchange interaction, and $D_x$ and $D_y$ are the anisotropy energies. For $D_x < D_y$, this system hosts low- and high-frequency magnon modes with eigenfrequencies of $\hbar\Omega_{l} = 2\sqrt{(J+D_y)D_x}$ and $\hbar\Omega_{h} = 2\sqrt{(J+D_x)D_y}$ \cite{Kampfrath2011,Rezende2019}. $Q_y$ and $Q_z$ are the phonon amplitudes of the two orthogonal components, which are taken to be circularly polarized in the $yz$-plane of the crystal, without loss of generality. $\mathbf{B}_{ph} = \mu_0 \gamma_{ph} \mathbf{L}_{ph}/V_c$ is the effective magnetic field produced by the chiral phonon mode, where $\gamma_{ph}$ is the gyromagnetic ratio of the phonon, $\mu_0$ is the vacuum permeability, and $V_c$ is the unit-cell volume. $\mathbf{L}_{ph}=\mathbf{Q}\times\dot{\mathbf{Q}}$ is the phonon angular momentum, where $\mathbf{Q}=(0,Q_y,Q_z)$. The magnetic moment of the magnon mode is given by $\mathbf{m}=\gamma_{el} \hbar(\mathbf{S}_1+\mathbf{S}_2)$, where $\gamma_{el}$ is the gyromagnetic ratio of the electron. Using these relations, Eq.~\eqref{eq:spin-phononAPPENDIX} can also be written as an angular momentum coupling, $V_{s\text{-}ph}=k \mathbf{L}_{ph}\cdot\mathbf{S}$, where $k=-\mu_0\gamma_{ph}\gamma_{el}\hbar/V_c$ is the angular momentum coupling coefficient. The circular polarization of the chiral phonon mode in the $yz$-plane ($\mathbf{L}_{ph}||\hat{\mathbf{x}}$) allows it to couple to the magnetic moment of the high-frequency magnon mode, $\mathbf{m}=(m_x,0,0)$, as illustrated in Fig.~\ref{fig:setup} in the main text. Finally, $\mathbf{p}_n=\mathbf{Z}_n Q_n$ is the electric dipole moment of component $n\in\{y,z\}$ of the chiral phonon mode, $\mathbf{Z}_n$ is its mode effective charge, and $\mathbf{E}(t)$ is the electric field component of the ultrashort laser pulse. The coordinate systems of the phonon eigenvectors can be chosen such that $n\equiv i$ for the components $Z_{n,i}$.

The coherent spin dynamics of the two sublattice spins, $\mathbf{S}_1$ and $\mathbf{S}_2$, are described by the Landau-Lifshitz-Gilbert equations \cite{Kampfrath2011,Rezende2019}, 

\begin{align}
    \frac{{\rm d}\mathbf{S}_1}{{\rm d}t} & =\frac{\gamma_{el}}{1+\kappa_{el}^2}\Big[\mathbf{S}_1 \times \mathbf{B}_1^{\rm eff} - \frac{\kappa_{el}}{|\mathbf{S}_1|} \mathbf{S}_1 \times (\mathbf{S}_1 \times \mathbf{B}_1^{\rm eff})\Big], \label{eq:LLG1APPENDIX}\\
    \frac{{\rm d}\mathbf{S}_2}{{\rm d}t} & =\frac{\gamma_{el}}{1+\kappa_{el}^2}\Big[\mathbf{S}_2 \times \mathbf{B}_2^{\rm eff} - \frac{\kappa_{el}}{|\mathbf{S}_2|} \mathbf{S}_2 \times (\mathbf{S}_2 \times \mathbf{B}_2^{\rm eff})\Big], \label{eq:LLG2APPENDIX}
\end{align}
where the lengths of the sublattice macrospins are normalized, $|\mathbf{S}_s|=1$ with $s\in\{1,2\}$, and where $\kappa_{el}$ is the Gilbert damping. The effective magnetic field acting on the spins can be derived from $\mathbf{B}^\mathrm{eff}_s = -(\gamma_{el}\hbar)^{-1}\partial_{\mathbf{S}_s} H$, where $H=H_0+V_{s\text{-}ph}$ is the total spin-dependent Hamiltonian, and is therefore given by

\begin{align}
    \mathbf{B}_1^\mathrm{eff} & = \mathbf{B}_{ph}-\frac{1}{\gamma_{el}}(J\mathbf{S}_{2}+2D_x S_{1,x}\hat{\mathbf{x}}+2D_y S_{1,y}\hat{\mathbf{y}}), \label{eq:magfield1APPENDIX}\\
    \mathbf{B}_2^\mathrm{eff} & = \mathbf{B}_{ph} -\frac{1}{\gamma_{el}}(J\mathbf{S}_{1}+2D_x S_{2,x}\hat{\mathbf{x}}+2D_y S_{2,y}\hat{\mathbf{y}}). \label{eq:magfield2APPENDIX}
\end{align}

At the same time, the coherent phonon dynamics are described by a phenomenological oscillator model \cite{Fechner2018,Juraschek2021}, $\ddot{Q}_n + \kappa \dot{Q}_n + \partial_{Q_n}H = 0$, which results in two equations of motions for the two components of the chiral phonon mode, $Q_y$ and $Q_z$,

\begin{align}
    \ddot{Q}_y & + \kappa_0 \dot{Q}_y + \Omega_0^2{Q}_y = Z_{y,i}  E_i(t) \nonumber\\
    & - k\left[ Q_{z}(\dot{S}_{1,x}+\dot{S}_{2,x})+2\dot{Q}_{z}(S_{1,x}+S_{2,x}) \right], \label{eq:phononeom1APPENDIX}\\
    \ddot{Q}_z & + \kappa_0 \dot{Q}_z + \Omega_0^2{Q}_z = Z_{z,i}  E_i(t) \nonumber\\
    & + k\left[ Q_{y}(\dot{S}_{1,x}+\dot{S}_{2,x})+2\dot{Q}_{y}(S_{1,x}+S_{2,x}) \right], \label{eq:phononeom2APPENDIX}
\end{align}
where $\kappa_0$ is the phonon linewidth.

The two components of the chiral phonon mode are driven by an ultrashort circularly polarized laser pulse and act as an effective magnetic field on the spins in the Landau-Lifshitz-Gilbert equations. In turn, the spins act as an additional driving force on the phonon modes in the phenomenological oscillator model. 

\subsection*{Quasistatic magnetization}

We now derive the implications this coupling has for the minimum position of the spin precession, by minimizing the Hamiltonian with respect to the spin vectors. We use the method of Lagrange multipliers to impose the constraint of spin normalization, $|\mathbf{S}_s|=1$,

\begin{align}
    \frac{\partial}{\partial \mathbf{S}_s} \big[ H & + \lambda_1(S_{1,x}^2+S_{1,y}^2+S_{1,z}^2-1) \nonumber\\
    & + \lambda_2(S_{2,x}^2+S_{2,y}^2+S_{2,z}^2-1) \big] = 0. \label{eq:minimizationAPPENDIX}
\end{align}
To obtain an analytical expression from the minimization procedure, we neglect the back-action of the spin system in the equations of motion \eqref{eq:phononeom1APPENDIX} and \eqref{eq:phononeom2APPENDIX} for the phonon modes, which makes the phonon angular momentum in Eq.~\eqref{eq:spin-phononAPPENDIX} spin independent. Eq.~\eqref{eq:minimizationAPPENDIX} therefore yields the following system of equations, 

\begin{align}
    J S_{2,x}+2(D_x-\lambda_1)S_{1,x}+kL_{ph,x}=0\label{eq:Sxcomponents1}, \\
    J S_{1,x}+2(D_x-\lambda_2)S_{2,x}+kL_{ph,x}=0\label{eq:Sxcomponents2}, \\
    J S_{2,y}+2(D_y-\lambda_1)S_{1,y}=0, \\
    J S_{1,y}+2(D_y-\lambda_2)S_{2,y}=0, \\
    J S_{2,z}-2\lambda_1 S_{1,z}=0 \label{eq:SZcomponents1}, \\ 
    J S_{1,z}-2\lambda_2 S_{2,z}=0 \label{eq:SZcomponents2}.
\end{align}
Furthermore, the constraint $S_{1z}+S_{2z}=0$ applies for the spin precession along the eigenvectors of the magnon modes, and from Eqs.~\eqref{eq:SZcomponents1} and \eqref{eq:SZcomponents2}, we obtain $\lambda_1=\lambda_2=-J/2$. Combining Eqs.~\eqref{eq:Sxcomponents1} and \eqref{eq:Sxcomponents2} therefore yields a nonzero minimum value for the magnetization, $M_{x,\mathrm{min}}=m_{x,\mathrm{min}}/V_c$, in the presence of a phonon angular momentum that is given by

\begin{align}
    M_{x,\mathrm{min}} &= \frac{\gamma_{el}\hbar}{V_c}(S_{1x}+S_{2x})_{\mathrm{min}} \nonumber\\ 
    &= -\frac{\gamma_{el}\hbar}{V_c}\frac{k}{J+D_x}L_{ph,x}.\label{eq:minimummagAPPENDIX}
\end{align}
This rectified magnetization, $\langle M_x\rangle \propto \langle L_{ph,x} \rangle \neq 0$, behaves similarly to the rectified phonon displacement in Eq.~\eqref{eq:phononminimumAPPENDIX}, where $\langle Q_c\rangle \propto \langle Q_d^2 \rangle \neq 0$. It corresponds to a quasistatic canting of the spins and the spin precession will follow around a transiently induced tilt angle, $\langle \theta \rangle \neq 0$. Eq.~\eqref{eq:minimummagAPPENDIX} is valid for the regime of small-angle deviations, $\theta\ll 90^\circ$, and breaks down for large $L_{ph,x}$.

\subsection*{Nonlinear magnetoelectric susceptibility}

Eq.~\eqref{eq:minimummagAPPENDIX} describes a quasistatic magnetization induced in the presence of a quasistatic phonon angular momentum. In order to describe the full response of the magnon mode to the effective magnetic field produced by the chiral phonon modes, we need to solve the Landau-Lifshitz-Gilbert equations. We will focus on the geometry shown in Fig.~\ref{fig:setup} for the high-frequency magnon mode in the system. The precession of the spins along the eigenvectors of this magnon mode fulfills the symmetry constraint $\textbf{S}_1 \equiv(S_x,S_y,S_z)$ and $\textbf{S}_2 \equiv(S_x,-S_y,-S_z)$. 

We can derive analytical expressions of the phonon amplitudes, $Q_y$ and $Q_z$, according to Eqs.~\eqref{eq:phononeom1APPENDIX} and \eqref{eq:phononeom2APPENDIX}. To first order in the electric field component of the laser pulse, the amplitudes are given in frequency space as
\begin{equation}
Q_y(\omega) = Z_{y}\frac{E_y(\omega)}{\Delta_0(\omega)}, ~~ Q_z(\omega) = Z_{z}\frac{E_z(\omega)}{\Delta_0(\omega)},
\end{equation}
where $\Delta_0(\omega)=\Omega_0^2-\omega^2+i\kappa_0\omega$. This results in an effective phono-magnetic field that is of second-order in the electric field and given in frequency space as
\begin{align}
    B_{ph,x}(\omega) &= \frac{\mu_0\gamma_{ph}}{V_c}L_{ph,x} \nonumber\\
    &= \frac{\mu_0\gamma_{ph}}{V_c}\epsilon_{xjk}Q_j(\omega) \circledast i\omega Q_k(\omega) \nonumber\\ 
    &= \frac{\mu_0\gamma_{ph}Z_jZ_k}{V_c}\epsilon_{xjk}\frac{E_j(\omega)}{\Delta_0(\omega)} \circledast \frac{i\omega E_k(\omega)}{\Delta_0(\omega)},    \label{eq:magneticfieldsolutionAPPENDIX}
\end{align}
where $\epsilon_{xjk}$ is the Levi-Civita tensor, $\circledast$ is the convolution operator, and $i,j,k \equiv x,y,z$.

With the effective phono-magnetic field acting as a driving force for the Landau-Lifshitz-Gilbert equations \eqref{eq:LLG1APPENDIX} and \eqref{eq:LLG2APPENDIX}, we can derive analytical approximations for the spin components $S_x$ and $S_y$. In linear order of $B_{ph,x}$, and therefore second order in the electric field component of the laser pulse, we obtain
\begin{align}
S_x(\omega) &= \frac{\gamma_{el}}{1+\kappa_{el}^2}\frac{i\kappa_{el}\omega+2 D_y/\hbar}{\Delta_m(\omega)}B_{ph,x}(\omega), \label{eq:Ssol1APPENDIX} \\
S_y(\omega) &= \frac{\kappa_{el}\gamma_{el}}{1+\kappa_{el}^2}\frac{i\omega}{\Delta_m(\omega)}B_{ph,x}(\omega).\label{eq:Ssol2APPENDIX}
\end{align}
Here, we have defined $\Delta_m(\omega)=\Omega_m^2-\omega^2+i\kappa_m\omega$, where $\Omega_m=\frac{2}{\hbar}\sqrt{\frac{(J+D_x)D_y}{1+\kappa_{el}^2}}$ is the damping-renormalized magnon frequency, and $\kappa_m=\frac{2\kappa_{el}}{\hbar\left(1+\kappa_{el}^2\right)}\left(J+D_x+D_y\right)$ is the magnon linewidth. These solutions return the expression for the rectified magnetization component in Eq.~\eqref{eq:minimummagAPPENDIX} when the frequency is set to zero, $\omega=0$. Using $M_x(0) = 2\gamma_{el}\hbar S_x(0)/V_c$, we obtain
\begin{align}
M_{x}(0) &= \frac{\gamma_{el}^2 \hbar^2}{V_c(J+D_x)}B_{ph,x}(0) \nonumber\\
 &= -\frac{\gamma_{el} \hbar}{V_c}\frac{k}{J+D_x}L_{ph,x}(0).
\end{align}

Using the full solution of $M_x(\omega) = 2\gamma_{el}\hbar S_x(\omega)/V_c$ in combination with Eq.~\eqref{eq:magneticfieldsolutionAPPENDIX}, we obtain

\begin{equation}
    M_x(\omega) = \frac{1}{\sqrt{2\pi}} \int_{-\infty}^{\infty}\chi^{(2)}_{me,xjk}(\omega,\omega')E_j(\omega-\omega')E_k(\omega')d\omega',
\end{equation}
where $\chi^{(2)}_{me,xjk}$ is the second-order nonlinear magnetoelectric susceptibility, given by 
\begin{align}
\chi^{(2)}_{me,xjk}(\omega,\omega') = & \frac{2\mu_0\gamma_{ph}\gamma_{el}^2\hbar\epsilon_{xjk} Z_yZ_z}{V_c^2\left(1+\kappa_{el}^2\right)} \nonumber\\
& \times \frac{\left(i\kappa_{el}\omega+2 D_y/\hbar\right)i\omega'}{\Delta_m(\omega)\Delta_0(\omega-\omega')\Delta_0(\omega')}.
\end{align}



%

\end{document}